\documentclass[twocolumn]{aastex631}

\usepackage{graphicx}
\usepackage{txfonts}
\usepackage{textcomp}
\usepackage{color}
\DeclareRobustCommand{\ion}[2]{%
\relax\ifmmode
\ifx\testbx\f@series
{\mathbf{#1\,\mathsc{#2}}}\else
{\mathrm{#1\,\mathsc{#2}}}\fi
\else\textup{#1\,{\mdseries\textsc{#2}}}%
\fi}

\def\softL{L\kern-0.8ex\raise0.1ex\hbox{'}\kern0.1ex}   
\def\gid{\mathrel{\mathchoice {\vcenter{\offinterlineskip\halign{\hfil
$\displaystyle##$\hfil\cr>\cr\noalign{\vskip1.2pt}=\cr}}}
{\vcenter{\offinterlineskip\halign{\hfil$\textstyle##$\hfil\cr>\cr
\noalign{\vskip1.2pt}=\cr}}}
{\vcenter{\offinterlineskip\halign{\hfil$\scriptstyle##$\hfil\cr>\cr
\noalign{\vskip1pt}=\cr}}}
{\vcenter{\offinterlineskip\halign{\hfil$\scriptscriptstyle##$\hfil\cr
>\cr
\noalign{\vskip0.9pt}=\cr}}}}}

\def\instituteA{Astronomical Institute, Slovak Academy of Sciences, 059 60 Tatransk\'a Lomnica, Slovakia}
\def\instituteB{Institute of Physics, Faculty of Science, Pavol Jozef \v{S}af\'arik University, Ko\v{s}ice, Slovakia}
\def\instituteC{Vysok\'e Tatry, 059 60 Tatransk\'a Lomnica 133, Slovakia}
\def\instituteD{Department of Theoretical Physics and Astrophysics, Masaryk University, Kotl\'a\v{r}sk\'a 2, 611 37 Brno, The Czech Republic}
\def\instituteE{MTA-ELTE Exoplanet Research Group, 9700 Szombathely, Szent Imre h. u. 112, Hungary}
\def\instituteF{ELTE Gothard Astrophysical Observatory, 9700 Szombathely, Szent Imre h. u. 112, Hungary}

\shorttitle{HD\,183986}
\shortauthors{Martin Va\v{n}ko et al.}

\begin{document} 

\title{HD\,183986: a high-contrast SB2 system with a pulsating component}
   
\author[0000-0002-2798-6944]{Martin Va\v{n}ko}
\affiliation{\instituteA}
\correspondingauthor{Martin Va\v{n}ko}
\email{vanko@astro.sk}

\author[0000-0003-3599-516X]{Theodor Pribulla}
\affiliation{\instituteA}

\author[0000-0003-1478-3256]{Pavol Gajdo\v{s}}
\affiliation{\instituteB}

\author[0000-0002-9125-7340]{J\'{a}n Budaj}
\affiliation{\instituteA}

\author{Juraj Zverko}
\affiliation{\instituteC}

\author[0000-0002-3304-5200]{Ernst Paunzen}
\affiliation{\instituteD}

\author[0000-0001-9483-2016]{Zolt\'{a}n Garai}
\affiliation{\instituteA}
\affiliation{\instituteE}
\affiliation{\instituteF}

\author[0000-0002-7204-9220]{{\softL}ubom\'{\i}r Hamb\'alek}
\affiliation{\instituteA}

\author[0000-0003-1890-3366]{Richard Kom\v{z}\'{\i}k}
\affiliation{\instituteA}

\author[0000-0003-4005-9245]{Emil Kundra}
\affiliation{\instituteA}
 
\begin{abstract}
{There is a small group of peculiar early-type stars on the main sequence that show 
different rotation velocities from different spectral lines. This inconsistency might be due
to the binary nature of these objects.}
{We aim to verify this hypothesis by a more detailed spectroscopic and photometric investigation of one such object:
HD\,183986.}
{We obtained 151 high and medium resolution spectra that covered an anticipated long orbital period.
There is clear evidence of the orbital motion of the primary component. We uncovered a very faint and broad spectrum of the 
secondary component. The corresponding SB2 orbital parameters, and the component spectra, were obtained by Fourier disentangling 
using the {\tt KOREL} code. The component spectra were further modeled by {\tt iSpec} code to arrive at the atmospheric quantities 
and the projected rotational velocities.}
{We have proven that this object is a binary star with the period 
$P=1268.2(11)$ d, eccentricity $e=0.5728(20)$, and mass ratio $q$ = 0.655. The primary component is a slowly rotating star 
(${v\sin i}$ = 27 km\,s$^{-1}$) while the cooler and less massive secondary rotates much faster ($v\sin i \sim$ 120 km\,s$^{-1}$). 
Photometric observations obtained by the \textit{TESS} satellite were also investigated to shed more light on this object. 
A multi-period photometric variability was detected in the \textit{TESS} data ranging from hours (the $\delta$ Sct-type variability) 
to a few days (spots/rotational variability). The physical parameters of the components and the origin of the photometric 
variability are discussed in more detail.}

\end{abstract}


%

\section{Introduction}

The easiest proof that a system is binary is the presence of eclipses. 
The second most useful technique is radial-velocity (RV) variations. 
Such variations are typically detected within spectroscopic surveys, which results in a list of objects cataloged 
according to their specific stellar attributes (e.g. \citealt{Duquennoy1991, Latham2002, Raghavan2010}). Once such
a variation has been detected, targeted and systematic observations must be obtained to characterize the system. In the simplest case, 
the binary comprises two components of relatively close luminosity which results in two systems of spectral lines in the composed 
spectrum. When the components differ significantly in luminosity, spectral lines of the fainter secondary component become less 
evident. The situation gets especially difficult when the secondary is a fast rotator. Although the spectral lines of the 
fast-rotating secondary component are shallow and hard both to identify and
to model, its light boosts the continuum level, which uniformly reduces the depths of the dominant component's spectral lines. 


More rarely, a star might be revealed to be double when discrepancies in the projected rotational 
velocity, ${v\sin i}$, determined using spectral lines that have a different sensitivity to the atmospheric effective 
temperature are detected, e.g., the lines of \ion{Mg}{ii} at $\lambda$ 4481 \AA~and \ion{Ca}{ii} $\lambda$ 3933 \AA\ \citep{Zverko2011}. 
\cite{Zverko2014} initiated a long-term spectroscopic survey of seven interesting CP candidates
thought to be binaries. 
Because the objects are bright, they are suitable for observations even with meter-class telescopes equipped 
with an \'echelle spectrograph, such as the 60-cm and 1.3-m reflectors of the Astronomical Institute, Slovak Academy of Sciences 
\citep{Vanko2020}. In this paper, we present a detailed study of HD\,183986, which was identified in this spectroscopic 
survey as a binary star candidate.

This paper is organized as follows. In Section~\ref{The star}, we present more details on HD\,183986. 
Section~\ref{spectro} describes new spectroscopic observations, the determination of RVs, the orbital
solution, and disentangling of the spectra. The modeling of spectra is shown in Section~\ref{iSpec}. The \textit{TESS} photometric 
data are analyzed and interpreted in Sections~\ref{tessphot}. An evolutionary state of HD\,183986 is studied in Section~\ref{evolution}. 
The paper is concluded with a discussion of the results and future observations which might improve the parameters of HD\,183986.

\section{HD\,183986}
\label{The star}

HD\,183986 ($V$ = 6.25, $\alpha_{2000}$ = 19$^{\rm h}$ 30$^{\rm m}$ 46.8$^{\rm s}$, 
$\delta_{2000}$ = +36$\degr$ 13$\arcmin$ 42$\arcsec$) is the brightest component of a visual triple star 
(see Table \ref{tab:star-params}). The B and C components are much fainter, 13.9 mag and 13.5 mag in $V$, and are 22.27~arcsec 
and 27.95~arcsec distant from component A, respectively \citep{Kuiper1961}. 

The duplicity of the brightest component was indicated per discrepant measurements of RV. \citet{Wenger2000} (Simbad database) list 
six different values of RV, ranging from 1.3 to 19 km\,s$^{-1}$. Unfortunately, the RV is mostly given as an average value from 
several measurements, without giving the individual values and the epochs of the observations. Measurements of the projected rotational 
velocity spread from 20 km\,s$^{-1}$ \citep{Abt2002} to 100 km\,s$^{-1}$ \citep{Palmer1968}. \citet{Hoffleit1995} determined 
${v\sin i}$ = 65 km\,s$^{-1}$ using the \ion{Ca}{ii} line $\lambda$ 3933 \AA, and \citet{Wolff1978} 
determined ${v\sin i}$ = 30 km\,s$^{-1}$ for \ion{Mg}{ii} $\lambda$ 4481 \AA. 

\begin{figure}[th]
\begin{center}
\includegraphics[width=\columnwidth]{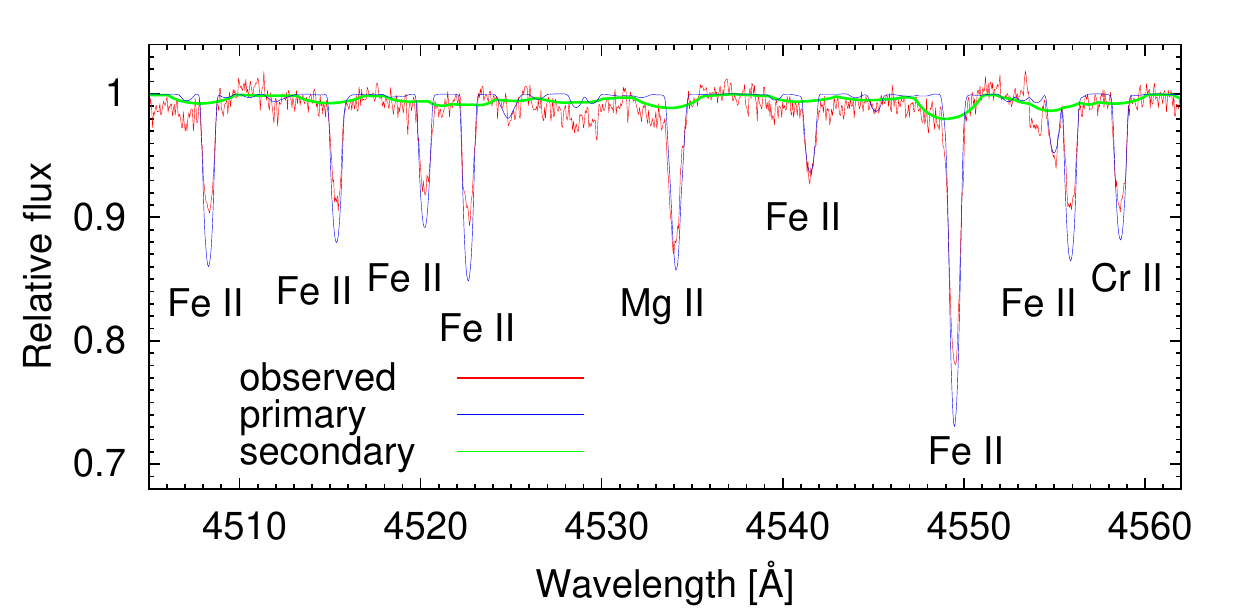}
\end{center}
\caption{A comparison of the observed spectrum with the synthetic ones of the primary and secondary. 
The latter is reduced by a factor of 0.255}
\label{fig:Reduced}
\end{figure}

The spectrum of HD\,183986 shows metalic lines reduced in strength by $\approx 20-25\%$ (Fig.~\ref{fig:Reduced}). 
It is possible that the object is a binary star, with the strength of the visible lines of the primary component diluted 
by the light of a second star with wide and hard-to-detect spectral features.

The astrometric solution presented in the Gaia EDR3 \citep{GaiaCollaborationDR3} gives the so-called astrometric over-noise 
parameter as large as $\sim$578 sigma. This indicates that the photocenter motion is marked and the Gaia DR3 will, very probably, 
provide the astrometric orbit of the system. The astrometric motion of the photocenter is also indicated by
the discrepant values of the parallax from the DR2 and EDR3 Gaia reductions (see Table~\ref{tab:star-params}).

\begin{table}[h]
\caption{Overview of identifications and basic parameters of HD\,183986. The references are given in the last column. 
The last five rows give Str\"omgren colour indices. The references are:
(1) - \citet{GaiaCollaborationDR2}, (2) - \citet{GaiaCollaborationDR3}, (3) - \citet{leeuwen}, (4) - \citet{hipp}, 
(5) - \citet{Wenger2000}, (6) - \citet{2MASS}, (7) - \citet{uvby}, (8) - \citet{Paunzen2015}}
\label{tab:star-params}
\begin{center}
\begin{tabular}{cccc}
\hline
\hline
	  BD     &                  &  +35\,3658  &       \\
          GSC    &                  &  02667-00744&       \\
          HIP    &                  &  95\,953      &     \\
\hline
 $\mu_{\alpha} \cos \delta $  &  [mas.yr$^{-1}$] &  1.158(106) &     (1) \\ 
 $\mu_{\delta}$  &  [mas.yr$^{-1}$] &  $-$12.114(109)&  (1) \\
          $\pi$  &     [mas]        &  4.561(63)     &  (1) \\  
          $\pi$  &     [mas]        &  5.007(69)     &  (2) \\  
          $\pi$  &     [mas]        &  5.18(45)      &  (3) \\
          $\pi$  &     [mas]        &  4.30(69)      &  (4) \\
\hline
   $V_{\rm max}$ &     [mag]        &  6.254         &  (2) \\
         $(B-V)$ &     [mag]        &  $-$0.018      &  (5) \\
         $J$     &     [mag]        &  6.194(24)     &  (6) \\
         $H$     &     [mag]        &  6.253(20)     &  (6) \\
         $K$     &     [mag]        &  6.250(16)     &  (6) \\ 
\hline
         $V$     &     [mag]        &  6.25          &  (7) \\   
         $b-y$   &     [mag]        &  0.002         &  (7) \\   
         $m_1$   &     [mag]        &  0.126         &  (7) \\   
         $c_1$   &     [mag]        &  0.958         &  (7) \\   
         $\beta$ &     [mag]        &  2.798(11)     &  (7) \\   
         $\beta$ &     [mag]        &  2.806(14)     &  (8) \\
\hline
\end{tabular}
\end{center}
\end{table}

\section{Spectroscopy}
\label{spectro}
\subsection{Observations and data reduction}

Spectroscopic observations of HD\,183986 were carried out with two different instruments. 
We began at the Star\'a Lesn\'a observatory (G1 pavilion) in July 2014 with a 60cm, f/12.5 Zeiss 
Cassegrain telescope equipped with a fiber-fed \'echelle spectrograph eShel \citep{Pribulla2015}. 
The spectra, consisting of 24 orders, cover the wavelength range from 4\,150 to 7\,600 \AA. 
The resolving power of the spectrograph is about $R =10\,000$. An Atik 460EX CCD camera, which has 
a 2749$\times$2199 array chip, 4.54$\mu$m square pixels, a read-out noise of 5.1 e$^{-}$ and
a gain of 0.26e$^{-}$/ADU, was used as a detector. From July 2017, we also observed at the Skalnat\'e Pleso 
Observatory (SP), using the 1.3m, f/8.36 Nasmyth-Cassegrain telescope, equipped with a fiber-fed \'echelle 
spectrograph of MUSICOS design \citep{BaudrandBohm1992}. The spectra were recorded using an Andor iKon-L DZ936N-BV 
CCD camera with a 2048$\times$2048 array, 13.5$\mu$m square pixels, 2.9e$^{-}$ read-out noise and
a gain close to unity. The spectral range of the instrument is 4\,250-7\,375 \AA~(56 \'echelle orders) with
a maximum resolution of $R=38\,000$. Our observations spanned an interval of 2\,334 days.

The raw data were reduced using {\tt IRAF} package tasks, Linux shell scripts, and {\tt FORTRAN} programs 
\citep{Pribulla2015,Garai2017}. In the first stage, master dark and flat-field frames were produced. 
In the second stage, photometric calibration of the frames was performed using dark and flat-field frames. 
Bad pixels were cleaned using a bad pixel mask and cosmic hits were removed using the \citet{Pych2004}
program. Usually, three consecutive photometrically-calibrated frames were combined to increase the signal-to-noise ratio 
(SNR) and to clean any remaining cosmics. The \'echelle order positions were defined by fitting the 6$^{th}$ order 
Chebyshev polynomials to a tungsten-lamp and blue LED spectra tracings. In the following stage, scattered light was modeled and 
subtracted. Aperture spectra were then extracted for both the object and the ThAr lamp frames and the resulting 2D spectra 
were dispersion solved. The 2D spectra were, lastly, combined into 1D spectra and rectified to the continuum.

\subsection{Primary component velocities and its orbit}
\label{SB1orbit}

\begin{figure}[h]
\begin{center}
\includegraphics[width=\columnwidth]{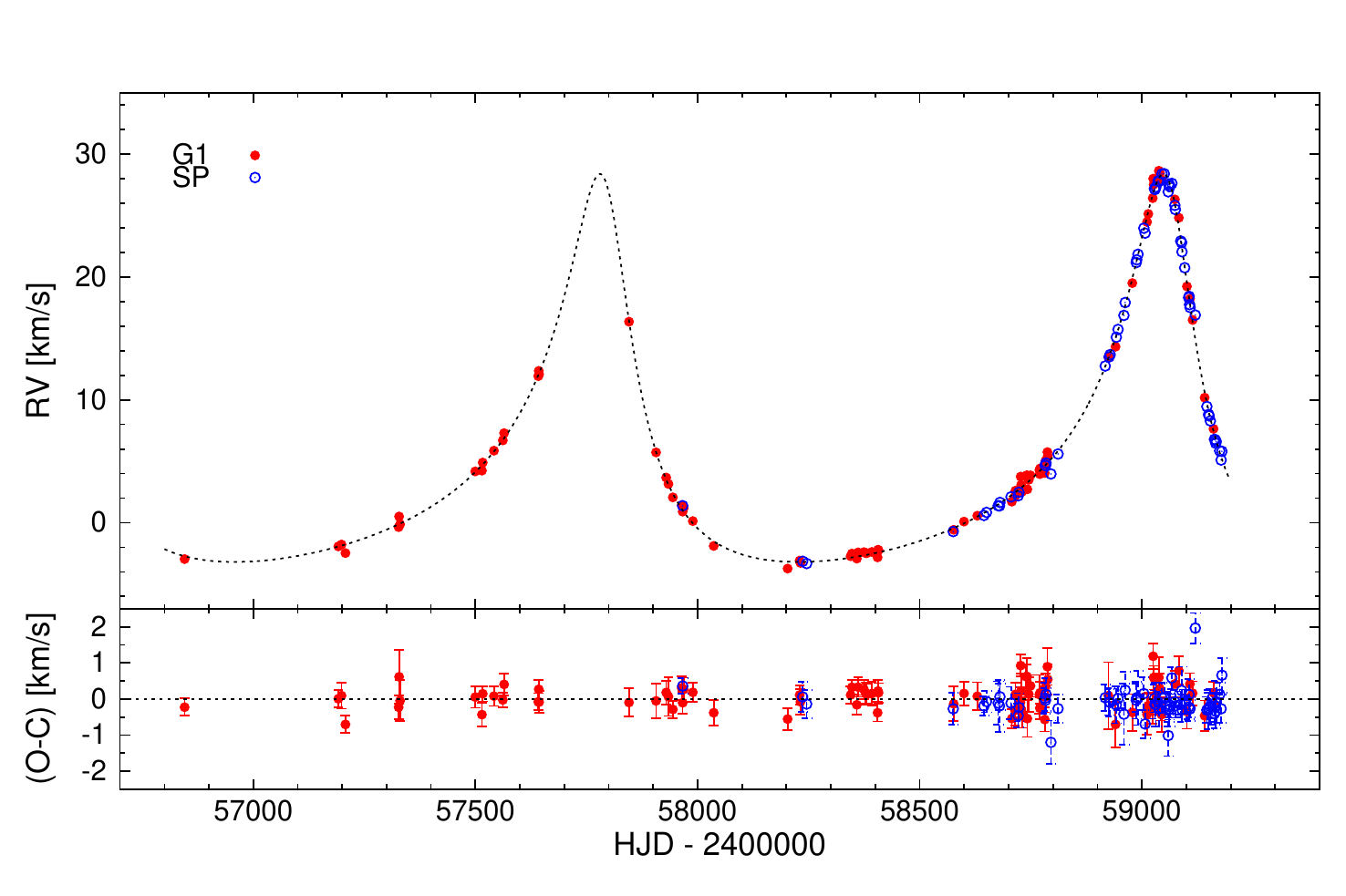}
\includegraphics[width=\columnwidth]{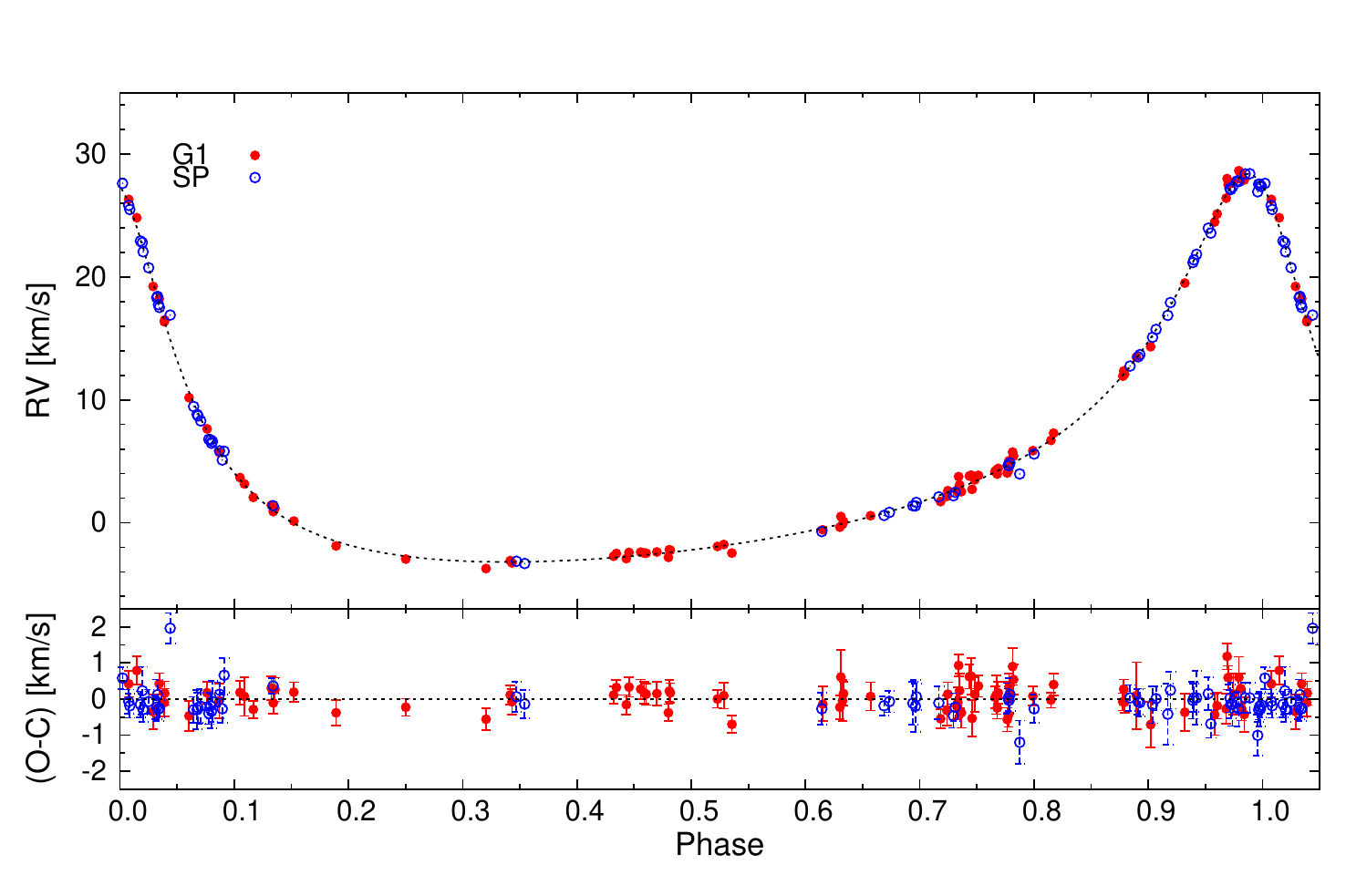}
\end{center}
\caption{Radial velocities of the primary component of HD\,183986 and their best fit and the residuals 
showing the measurement errors. Radial velocity as a function of time (top) and the corresponding phase 
diagram (bottom) is shown. Data obtained in the G1 pavilion with the eShel spectrograph
(G1), together with those obtained with the MUSICOS at the Skalnat\'e Pleso observatory (SP), are shown.}
\label{fig:SB1}
\end{figure}

The RVs were determined using the cross-correlation function (CCF) technique \citep{griffin67, simkin74, tonry1979, Zverko2007}. 
Except for the Balmer lines, the observed spectrum of the star included a few tens of metallic absorption lines that
were largely of central depths $\approx 0.1$ below the continuum. The strongest one
was the line of \ion{Mg}{ii} $\lambda$ 4481 \AA\, with a central 
depth of $\approx0.45$ in the spectra, with $R$ = 38\,000. We used two wavelength regions, namely 4\,400--4\,640~\AA\ 
and 4\,900--5\,470 \AA, which contain an absolute majority of the mentioned metallic lines. 
Due to the luminosity ratio and the high $v \sin i$, which was estimated to be up to 150 km\,s$^{-1}$, the CCFs represent only the 
primary component and its orbital motion. The synthetic ''template'' spectrum used here was computed 
with $T_{\rm eff} =10\,600$~ K and $\log g=3.63$ rotated with $v \sin i = 30$ km\,s$^{-1}$, adopted from \citet{Zverko2013}. 
The RVs are listed in Tables \ref{tab:radialG1} and \ref{tab:radialSP} and plotted in Fig.~\ref{fig:SB1}.

Having a sufficient number of spectra for each spectrograph (eShel and MUSICOS), the RV errors were first 
determined as 1/SNR \citep[see equation 1 of][]{Hatzes2010}. These errors were then re-scaled to give the 
reduced $\chi^2$ = 1 for either of the spectrographs. The scaling constant in $\sigma = C/SNR$ was found to 
be 28.6 for MUSICOS and 31.4 for eShel. No statistically significant shift of the systemic RV between the 
spectrographs was found. 

All the RVs were modeled simultaneously to obtain the spectroscopic elements. The modeling was performed 
using a gradient-based optimization (GbO). The best fit to the data showed three deviating points at 
HJD 2\,458\,343.51, 2\,458\,347.38 and 2\,458\,358.45, which were omitted, and
the RVs were re-analyzed. 
The resulting orbital parameters for the primary component are listed in
Table~\ref{SB1}, and the best fit with corresponding residuals in Fig.~\ref{fig:SB1}. 
The mean standard deviation of a single RV measurement 
is about 0.36 km\,s$^{-1}$. This is comparable to the systematic uncertainties caused by the limited stability 
of the spectrographs. Using the zero point difference of the wavelength solution of the preceding and following 
comparison spectra, the typical RV stability is 0.20 km\,s$^{-1}$, and 0.05 km\,s$^{-1}$ for MUSICOS and eShel 
spectrographs, respectively.

We also employed an alternative {\tt FOTEL} code \citep{Hadrava1990}, which enables either a simultaneous or 
a separate analysis of the light curve and radial velocity of a binary star. It is based on the minimization of 
the sum $\Sigma(O-C)^2$ as a function of the orbital elements. The orbital elements of the primary component's orbit, 
obtained with the same RV data as in the GbO modelling, are listed in the last column of Table~\ref{SB1}. 
The resulting parameters are mostly within errors of the 
GbO results.

\begin{table}[h]
\caption{Spectroscopic orbit of the primary component obtained fitting radial velocities 
from the CCF method. The standard error of the last decimal place is given in parenthesis.}
\label{SB1}
\begin{center}
\begin{tabular}{llcc}
\hline
\hline
Element &                 &   GbO            &  FOTEL  \\
\hline
$P$     & [days]          & 1\,268.2(11)     &  1\,267.8(12)   \\
$e$     &                 & 0.572\,8(20)     &  0.571\,4(22) \\
$\omega$& [rad]           & 0.354(6)         &  0.359(7)   \\
$T_0$   & [HJD]           & 2\,456\,528.2(24)&  2\,456\,529.4(27) \\
$V_0$   & [km\,s$^{-1}$]  & 4.115(32)        &  4.030(40)      \\
$K_1$   & [km\,s$^{-1}$]  & 15.79(5)         &  15.83(7) \\ 
\hline
$a_1 \sin i$ & [a.u.]     & 1.509(5)         &  1.514(7) \\
$f(m)$     & [M$_\odot$]  & 0.286(3)         &  0.288(4) \\
$\chi^2$   &              & 147.64           &     -     \\
d.o.f.     &              & 148-6            &     -     \\
$\Sigma(O-C)^2$ &         &                  &  0.0660   \\ 
\hline
\end{tabular}
\end{center}
\end{table}

\subsection{Signatures of the secondary component}
\cite{Zverko2013} demonstrated that the observed line profiles of the 
\ion{Ca}{ii} $\lambda$ 3933 \AA\ and \ion{Mg}{ii} $\lambda$ 4481 \AA\ lines of HD\,183986 can 
be reproduced by superimposing the spectra of two stars, namely a B9.5III and a middle A-type star. 
They also estimated the flux ratio of the components as $F_1/F_2$ = 0.9/0.1 in the ultraviolet 
region and $F_1/F_2$ = 0.89/0.11 in the blue region.

Now, having the orbital cycle sufficiently covered by the RV observations, we can illustrate how, 
for example, the profile of the line of \ion{Mg}{ii} $\lambda$ 4481 \AA\ varies due to the orbital motion. 
The spectral lines of the secondary component were recognized by a careful inspection of the spectrum. 
They are very weak due to the luminosity ratio of the component stars, and in addition, the lines are broadened 
due to the high projected rotational velocity, $v \sin i$, which may reach up to $\approx$ 150 km\,s$^{-1}$.
 
\begin{figure}[h]
\begin{center}
\includegraphics[width=\columnwidth]{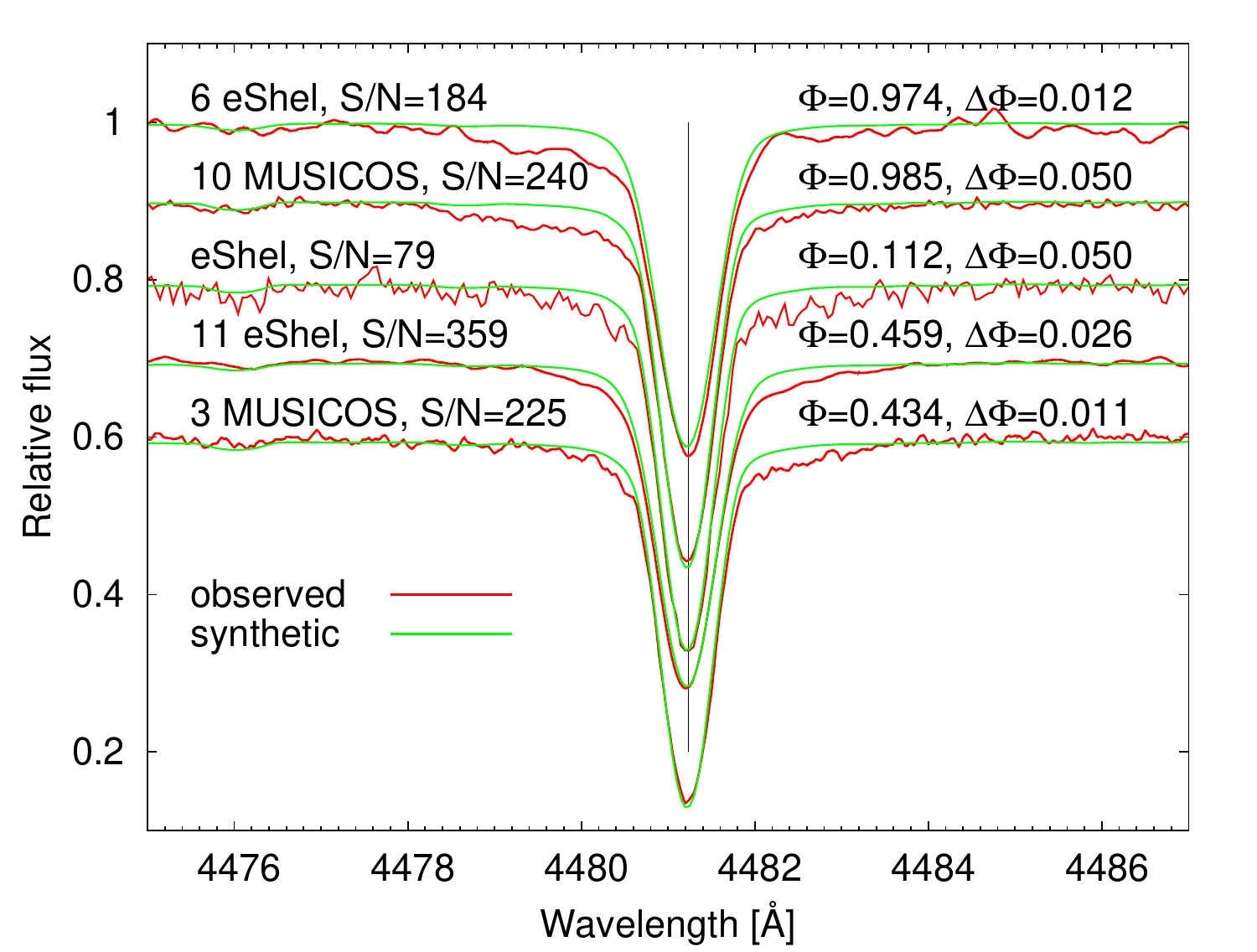}
\end{center}
\caption{The profile of the \ion{Mg}{ii} $\lambda$ 4481 \AA\ near the maximum RV of the primary component (upper 2 spectra), 
close to the systemic velocity (middle), and the minimum RV (lower 2 spectra). 
The labels on the left mean that 6, 10, 11 and 3 individual spectra have been co-added to increase the SNR. 
The corresponding mean phase, $\Phi$, and the phase interval, $\Delta \Phi$, are shown on the right.}
\label{fig:T_MIN_MAX}
\end{figure}

\begin{figure}[h]
\begin{center}
\includegraphics[width=\columnwidth]{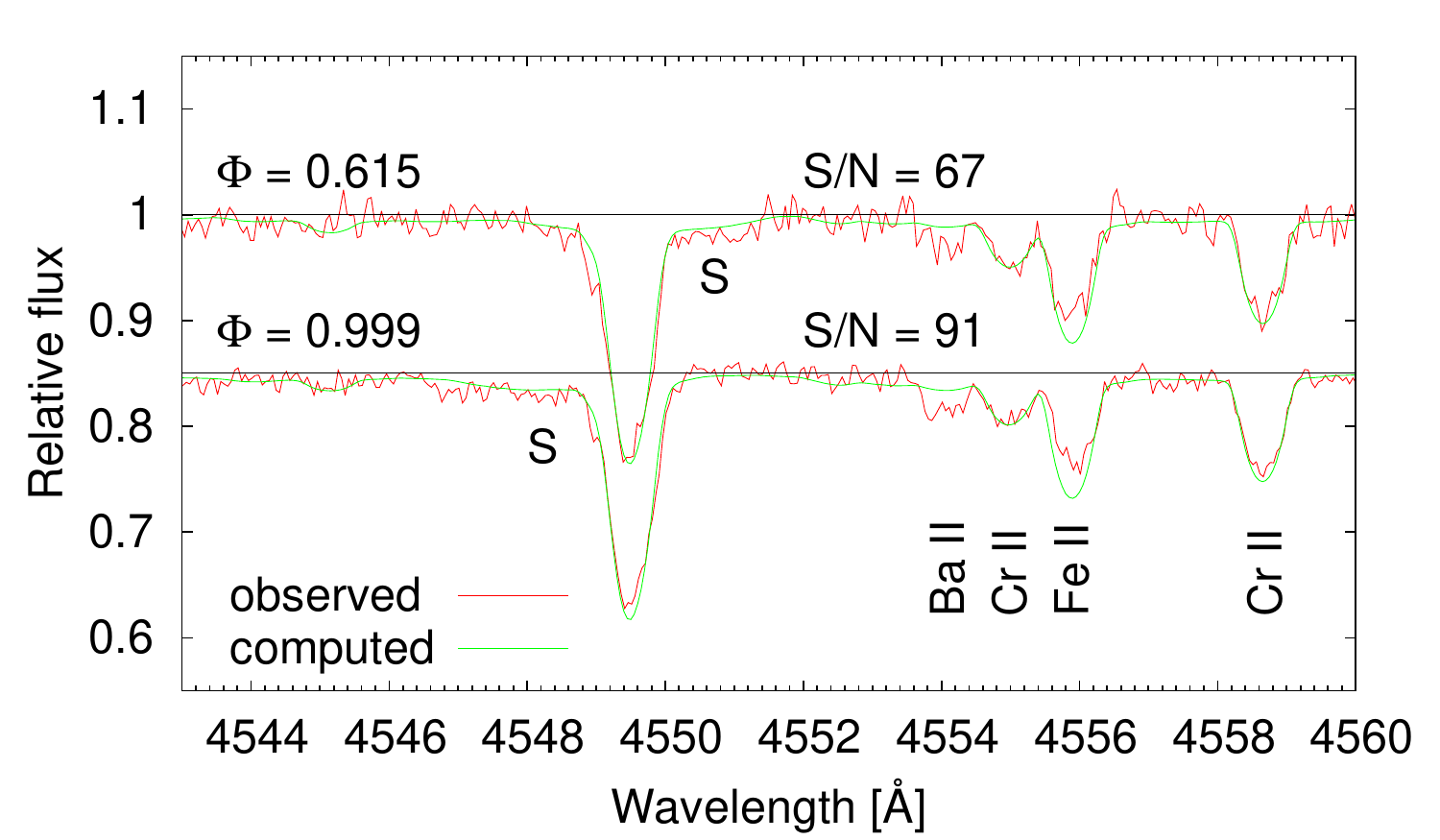}
\end{center}
\caption{Similar to Fig.~\ref{fig:T_MIN_MAX}, but in the vicinity of the \ion{Fe}{ii} line $\lambda$ 4549 \AA. 
The computed spectrum is a theoretical one summed from the spectra of the primary and secondary components. 
The letter "S" marks the position of the secondary component line.}
\label{fig:FeII_4549}
\end{figure}

Spectra selected close to the extremes of the RV curve, as well as near the systemic
velocity, are plotted 
in Fig.~\ref{fig:T_MIN_MAX}. It is well documented that, when the secondary component approaches, a depression 
in the short-wavelength wing of the line appears while, when it recedes, depression occurs in the 
red wing of the line. A similar depression in the opposite wings of the line can be clearly seen in the 
line of the \ion{Fe}{ii} $\lambda$ 4549 \AA\ (Fig.~\ref{fig:FeII_4549}). 
The line is, in terms of strength, the second following the line of magnesium \ion{Mg}{ii} $\lambda$ 4481 \AA. 
The theoretical spectrum shown here was computed assuming the atmospheric parameters adopted from \cite{Zverko2013}, 
namely $T_{\rm eff}=10\,600$~K, $\log g=3.63$ for the primary and $T_{\rm eff}=8\,200$~ K, $\log g=4.2$ 
for the secondary, and taking into account the ratio of luminosities of the components.


\begin{table}[h]
\caption{Spectroscopic SB2 orbit obtained by the Fourier disentangling. 
The standard error of the last decimal place is given in parenthesis. 
{\tt KOREL} does not calculate errors of parameters. 
The errors of $V_0, K_1, K_2, a_{1,2} \sin i, M_{1,2} \sin^3 i$ and $\Sigma(O-C)^2$ are from {\tt FOTEL}}
\label{SB2}
\begin{center}
\begin{tabular}{lcc}
\hline
\hline
Parameter       &                          &                \\
\hline
$P$             & [days]                   &  1268.2        \\
$e$             &                          &  0.5729        \\
$\omega$        & [rad]                    &  0.359         \\
$T_0$           & [HJD]                    &  2\,456\,528.2 \\
$V_0$           & [km\,s$^{-1}$]           &  4.03(4)       \\
$K_1$           & [km\,s$^{-1}$]           & 15.82(5)       \\
$K_2$           & [km\,s$^{-1}$]           & 24.15(5)       \\
\hline
$a_1 \sin i$    & [a.u.]                   &   1.514\,4     \\
$a_2 \sin i$    & [a.u.]                   &   2.247\,5     \\
$M_1 \sin^3 i$  & [M$_\odot$]              &   2.93         \\
$M_2 \sin^3 i$  & [M$_\odot$]              &   1.92         \\
$q$             &                          &   0.655        \\                                                                     
$\Sigma(O-C)^{\rm 2}$ &                    &   0.138\,1     \\
\hline
\end{tabular}
\end{center}
\end{table}

\begin{deluxetable*}{cDrcDrcDr}
\tablecaption{Heliocentric radial velocities of HD\,183986 from G1. The estimated SNRs at 4400 \AA~ are also listed. 
\label{tab:radialG1}}
\tablewidth{0pt}
\decimals
\tablehead{
\colhead{HJD} & \multicolumn2c{RV} & \colhead{SNR} & \colhead{HJD} & \multicolumn2c{RV} & \colhead{SNR} & \colhead{HJD} & \multicolumn2c{RV} & \colhead{SNR} \\
\colhead{2\,400\,000+} & \multicolumn2c{[km\,s$^{-1}$]} & \colhead{} & \colhead{2\,400\,000+} & \multicolumn2c{[km\,s$^{-1}$]} & \colhead{} & \colhead{2\,400\,000+} & \multicolumn2c{[km\,s$^{-1}$]} & \colhead{}}
\startdata
 56845.4475 & -2.95 & 132 & 58231.6029 & -3.27 &  82 & 58769.2555 &  4.30 &  55\\
 57191.5324 & -1.93 & 126 & 58344.3883 & -2.73 & 132 & 58770.2446 &  3.96 & 105\\
 57198.4525 & -1.77 &  86 & 58347.3794 & -2.50 & 134 & 58771.2281 &  4.42 &  99\\
 57207.4420 & -2.47 & 128 & 58358.3795 & -2.92 & 106 & 58781.2150 &  4.05 &  94\\
 57327.2816 & -0.35 &  87 & 58361.3937 & -2.41 & 107 & 58782.2516 &  4.71 &  74\\
 57328.3314 &  0.51 &  40 & 58374.3454 & -2.39 & 109 & 58783.2421 &  5.04 &  82\\
 57330.3210 & -0.12 &  53 & 58378.3288 & -2.46 &  98 & 58787.2984 &  5.75 &  59\\
 57499.5912 &  4.18 &  98 & 58380.3630 & -2.48 & 105 & 58788.2476 &  5.43 &  70\\
 57514.5199 &  4.24 &  94 & 58392.3203 & -2.38 &  94 & 58924.6172 & 13.48 &  32\\
 57516.5285 &  4.90 & 138 & 58405.2533 & -2.81 & 123 & 58940.5006 & 14.33 &  48\\
 57541.5202 &  5.87 & 116 & 58406.2512 & -2.20 & 100 & 58978.3835 & 19.50 &  58\\
 57561.5570 &  6.71 & 144 & 58407.2656 & -2.24 & 108 & 59011.4915 & 24.48 &  53\\
 57564.5105 &  7.29 & 101 & 58576.5496 & -0.57 &  62 & 59014.3394 & 25.14 &  81\\
 57641.3648 & 11.93 & 119 & 58599.5643 &  0.10 &  92 & 59024.3419 & 26.42 &  73\\
 57642.3621 & 12.36 & 116 & 58629.5304 &  0.57 &  75 & 59025.4075 & 28.00 &  83\\
 57643.3026 & 12.09 & 104 & 58707.3115 &  1.72 & 115 & 59026.4321 & 27.53 &  90\\
 57845.5650 & 16.36 &  77 & 58714.3568 &  2.14 &  73 & 59032.3348 & 27.70 &  54\\
 57906.5262 &  5.72 &  62 & 58715.3577 &  2.61 &  91 & 59038.3680 & 28.65 &  52\\
 57929.5448 &  3.68 &  93 & 58725.3420 &  2.56 &  84 & 59040.4144 & 28.46 &  70\\
 57934.5240 &  3.16 &  56 & 58726.4111 &  2.70 &  71 & 59041.4208 & 27.95 &  91\\
 57944.4365 &  2.07 & 126 & 58727.3460 &  3.75 &  98 & 59044.4157 & 27.90 &  63\\
 57964.4208 &  1.43 &  90 & 58728.3038 &  3.08 &  63 & 59074.3292 & 26.32 &  83\\
 57966.4706 &  0.91 &  98 & 58730.3559 &  2.53 &  74 & 59083.2849 & 24.82 &  79\\
 57968.4001 &  1.18 &  93 & 58739.2977 &  3.80 &  91 & 59101.2839 & 19.23 &  62\\
 57989.4201 &  0.14 & 116 & 58741.2609 &  3.87 &  59 & 59108.2470 & 18.23 &  99\\
 58036.2989 & -1.89 &  84 & 58742.2569 &  2.73 &  59 & 59114.2379 & 16.50 &  95\\
 58202.6364 & -3.73 &  98 & 58745.2722 &  3.50 &  62 & 59141.1980 & 10.18 &  70\\
 58229.6100 & -3.08 & 116 & 58749.2397 &  3.86 &  98 & 59161.2816 &  7.64 & 104\\
\enddata
\end{deluxetable*}

\begin{deluxetable*}{cDrcDrcDr}
\tablecaption{Heliocentric radial velocities of HD 183986 from SP. Three systematically deviating 
points, 2 458 343.50870, 2 458 347.38010 and 2 458 358.44960, were not used for the next analysis. 
The estimated SNRs at 4400 \AA~ are also listed.
\label{tab:radialSP}}
\tablewidth{0pt}
\decimals
\tablehead{
\colhead{HJD} & \multicolumn2c{RV} & \colhead{SNR} & \colhead{HJD} & \multicolumn2c{RV} & \colhead{SNR} & \colhead{HJD} & \multicolumn2c{RV} & \colhead{SNR} \\
\colhead{2\,400\,000+} & \multicolumn2c{[km\,s$^{-1}$]} & \colhead{} & \colhead{2\,400\,000+} & \multicolumn2c{[km\,s$^{-1}$]} & \colhead{} & \colhead{2\,400\,000+} & \multicolumn2c{[km\,s$^{-1}$]} & \colhead{}}
\startdata
 57966.3834 &  1.38 & 144 & 58942.5265 & 15.10 &  59 & 59075.4500 & 25.50 &  89\\
 58236.5091 & -3.13 &  69 & 58946.5072 & 15.73 &  89 & 59087.3447 & 22.93 &  84\\
 58245.5391 & -3.32 &  79 & 58959.5131 & 16.88 &  36 & 59089.4625 & 22.80 &  45\\
 58343.5087 & -4.65 &  83 & 58962.4249 & 17.92 &  59 & 59090.3030 & 22.06 &  92\\
 58347.3801 & -4.33 & 112 & 58987.4497 & 21.18 &  59 & 59096.3852 & 20.76 &  88\\
 58358.4496 & -4.65 &  88 & 58988.4476 & 21.40 &  76 & 59105.3386 & 18.32 &  93\\
 58575.4717 & -0.72 &  67 & 58991.4351 & 21.85 &  40 & 59106.2910 & 18.43 &  73\\
 58644.4567 &  0.60 & 109 & 59004.4100 & 23.98 &  67 & 59107.3438 & 17.76 &  95\\
 58650.4295 &  0.85 &  96 & 59007.3186 & 23.58 &  75 & 59108.3786 & 17.52 & 119\\
 58676.4817 &  1.38 &  51 & 59028.4739 & 27.19 &  79 & 59120.2311 & 16.90 &  70\\
 58679.4284 &  1.37 &  41 & 59029.4058 & 27.12 &  75 & 59146.2498 &  9.46 &  86\\
 58680.3978 &  1.65 &  86 & 59031.3720 & 27.32 &  50 & 59150.2405 &  8.81 &  57\\
 58705.4143 &  2.10 &  67 & 59036.4256 & 27.83 & 106 & 59151.2793 &  8.70 &  69\\
 58721.4782 &  2.19 &  92 & 59038.4410 & 27.78 &  62 & 59154.2333 &  8.29 &  63\\
 58723.4627 &  2.48 &  79 & 59045.3634 & 28.39 &  73 & 59163.2130 &  6.80 &  98\\
 58782.2594 &  4.68 &  71 & 59050.4923 & 28.41 &  72 & 59165.3082 &  6.74 &  86\\
 58783.2614 &  4.66 &  91 & 59059.3083 & 26.94 &  52 & 59166.1999 &  6.47 &  63\\
 58784.2637 &  4.86 &  66 & 59060.3910 & 27.56 &  89 & 59167.2362 &  6.61 &  54\\
 58795.2202 &  3.97 &  51 & 59061.4166 & 27.55 &  85 & 59175.1781 &  5.85 &  93\\
 58811.1984 &  5.59 &  76 & 59062.3655 & 27.39 &  64 & 59178.2052 &  5.11 &  79\\
 58917.6143 & 12.76 &  83 & 59063.3634 & 27.39 &  91 & 59180.2369 &  5.82 &  62\\
 58926.5886 & 13.50 &  83 & 59067.4542 & 27.62 &  99 &     .      &   .   &  . \\
 58928.5464 & 13.68 &  78 & 59074.3237 & 25.83 &  90 &     .      &   .   &  . \\
\enddata
\end{deluxetable*}

\begin{figure*}[th]
\begin{center}
\includegraphics[width=17cm]{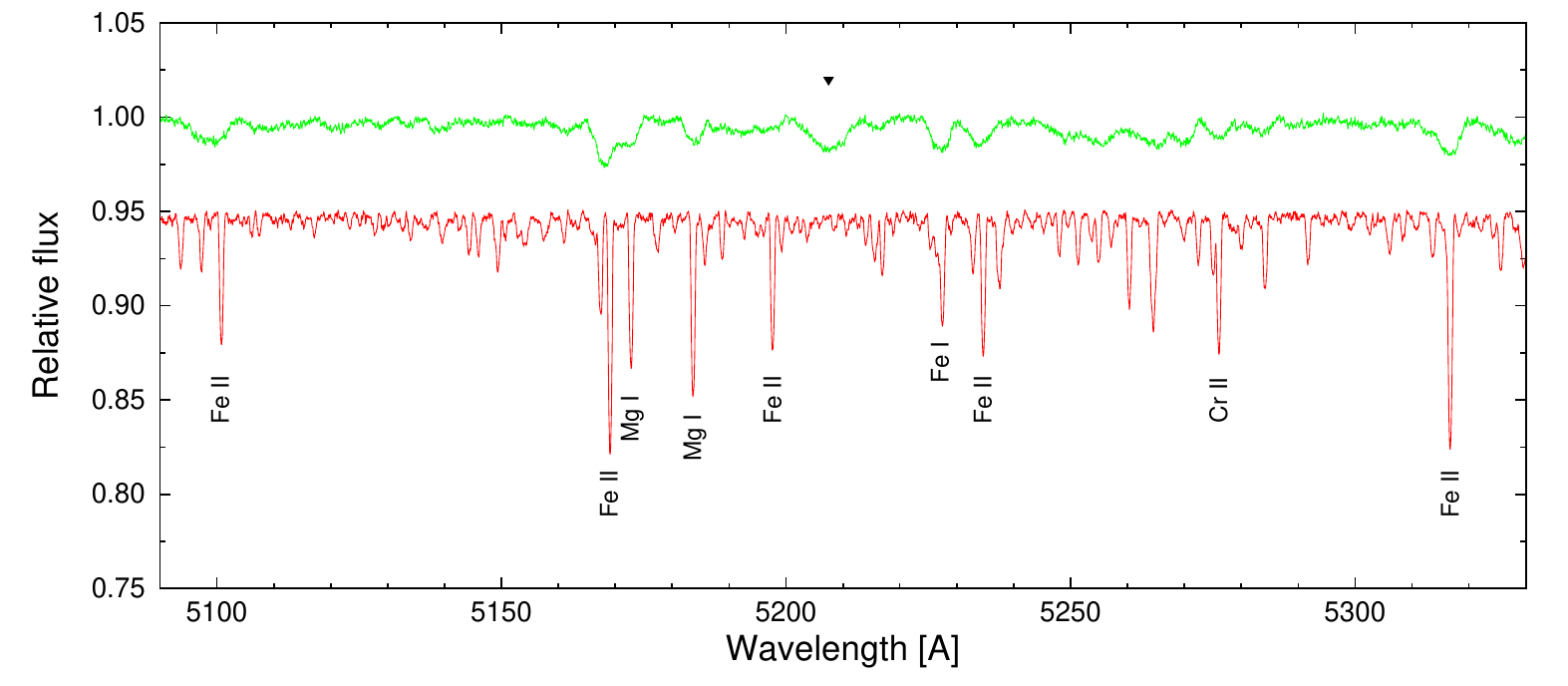}
\end{center}
\caption{The disentangled spectra in the yellow region normalised to unity. 
Some of the spectral lines identified in the spectrum of the primary component 
(lower spectrum) are also identifiable as rotationally-widened in the spectrum 
of the cooler secondary component (upper spectrum). A remaining artifact
depression, resulting from the disentangling, is visible in the secondary-spectrum (black triangle). 
The primary component spectrum was shifted for clarity by $-$0.05.}
\label{fig:Rozpletene}
\end{figure*}

\subsection{Disentangling spectra with KOREL}
\label{korel}

Spectral disentangling is a powerful method used to separate component spectra of spectroscopic binaries (SB2) 
and multiple systems. There are several viable approaches, e.g. disentangling in the wavelength domain 
\citep{Simon1994,Ilijic2004}, Fourier disentangling \citep{Hadrava1995}, or tomographic decomposition 
\citep{Bagnuolo1991, Konacki2010}. Each of these techniques has different assumptions. While some of them 
require knowledge of radial velocities, or of basic orbital parameters, others provide results 
with a complete solution without RVs as an input. Nevertheless, spectral disentangling is the method of 
choice for high-contrast systems. The disentangling typically provides spectra of the fainter components 
with a higher SNR, which can be used for spectroscopic analysis. The advantages and limitations of the above 
techniques are discussed in detail by e.g., \citet{Ilijic2004} or \citet{Helminiak2019}. 

In order to separate the component spectra of HD\,183986, the Fourier disentangling method of \citet{Hadrava1995} 
was used. {\tt KOREL} \citep{Hadrava2004} is a code for decomposition of component spectra and determination 
of the orbital elements of binary and multiple stars by means of Fourier disentangling. It is especially suitable for 
spectra with faint spectral lines of the secondary component superimposed on strong lines of primary ones. 
It enables multiple spectral regions to be analyzed simultaneously in a set of spectra. 

We selected 23 out of 67 MUSICOS (SP) spectra, 15 of them with SNR$\gid 89$; a further 8 with SNR$\gid 62$ were added 
for a better covering of the orbital phase. Two sections well populated with metallic lines were selected, namely 
from 4\,377--4\,650~\AA~and 5\,031--5\,359~\AA, avoiding the region of the $H_\beta$ line, where the continuum position 
is often uncertain. The wavelength scale was converted to the RV scale with a constant step of 4.3
km\,s$^{-1}$, thus giving 4\,096 = $2^{12}$ frequency points. 

Having an initial set of orbital elements, the code enables each parameter, individually, to be adjusted while the others 
are kept fixed. Furthermore, all six elements can be adjusted simultaneously. Various combinations
of fixed and adjusted elements are also possible.   

We adopted the elements corresponding to the orbit of the primary component that were derived using both the 
eShel and MUSICOS data as they together span over 2\,334 days, while the MUSICOS data used in {\tt KOREL} covers 
only 1\,214 days, which is slightly lower than the orbital period.

First, we fixed the orbital period and the time of the periastron passage, and
the eccentricity was adjusted while keeping the remaining elements fixed. We then adjusted the argument of periastron $\omega$, 
the semi-amplitude $K_1$, and the mass ratio $q=M_2/M_1$ with the new values of the preceding elements fixed. 
Next, those four elements were converged simultaneously. And lastly, all the 6 elements were converged simultaneously. 
The resulting parameter set is given in Table~\ref{SB2}. The disentangled spectra are plotted 
in Fig.~\ref{fig:Rozpletene}. In the case of the secondary component, some artifacts very probably resulting from the 
disentangling are visible.

\section{Spectra modeling}
\label{iSpec}

The {\tt KOREL} disentangling does not provide the flux ratio of the components \citep[see e.g.,][]{renormal2,renormal}. Hence,
prior to further modeling, the component spectra must be corrected for the contribution of the other component. 
We can estimate the flux ratio from the observed color of the system and the mass ratio determined by the disentangling. 
Applying the programs {\tt UVBYBETA} \citep{Moon1985} and {\tt TEFFLOG}
\citep{Moon1985}, using the $uvby\beta$ indices of HD\,183986
\citep{Wenger2000}, we get $T_{\rm eff}=10\,490$~K. As initial values, we used the data summarized in the 
Astrophysical Quantities (TAQ) \citep{Cox2000}. This temperature corresponds to a B9.5V star with $M_1 = 3.1$ M$_\sun$ 
and $R_1$ = 2.70 R$_\sun$. Adopting the {\tt KOREL} mass ratio, $q = 0.655$, the secondary mass is about $M_2 = 2.0$ M$_\sun$. 
This corresponds to an A5V star with about $R_2$ = 1.70 R$_\sun$. Now we can estimate the flux ratio of the components in both 
the blue and yellow regions. In Tab. 15.8 of TAQ, we estimate B9.5V, $M(V)=0.425$\,mag for the primary, and with a corresponding 
$B-V=-0.045$\,mag, we get $M(B)=0.380$\,mag. For the secondary A5V, there is $M(V)=1.95$\,mag, $B-V=0.15$\,mag, 
and $M(B)=2.10$\,mag. Using $M(B)_{\rm prim}-M(B)_{\rm sec}=-2.5\times\log(F_1/F_2)$, we arrive at
a flux ratio of $F_2/F_1 = 0.17/0.83$ = 0.205 in the $B$ band. For the
$V$ band, we get $F_2/F_1 = 0.20/0.80$ = 0.245.

The component spectra were modeled to determine the atmospheric parameters $T_{\rm eff}$, $\log g$, metallicity, 
and the projected rotational velocity, $v \sin i$.  We used code {\tt iSpec} \citep{blanco14,blanco19} based 
on the {\tt SPECTRUM} \citep{spectrum}. Prior to the modeling, the line depth of either component was corrected 
(increased) to take into account the contribution of the other component. The (continuum) flux ratios estimated 
above were taken into account. 

\begin{figure}[h]
\begin{center}
\includegraphics[width=\columnwidth]{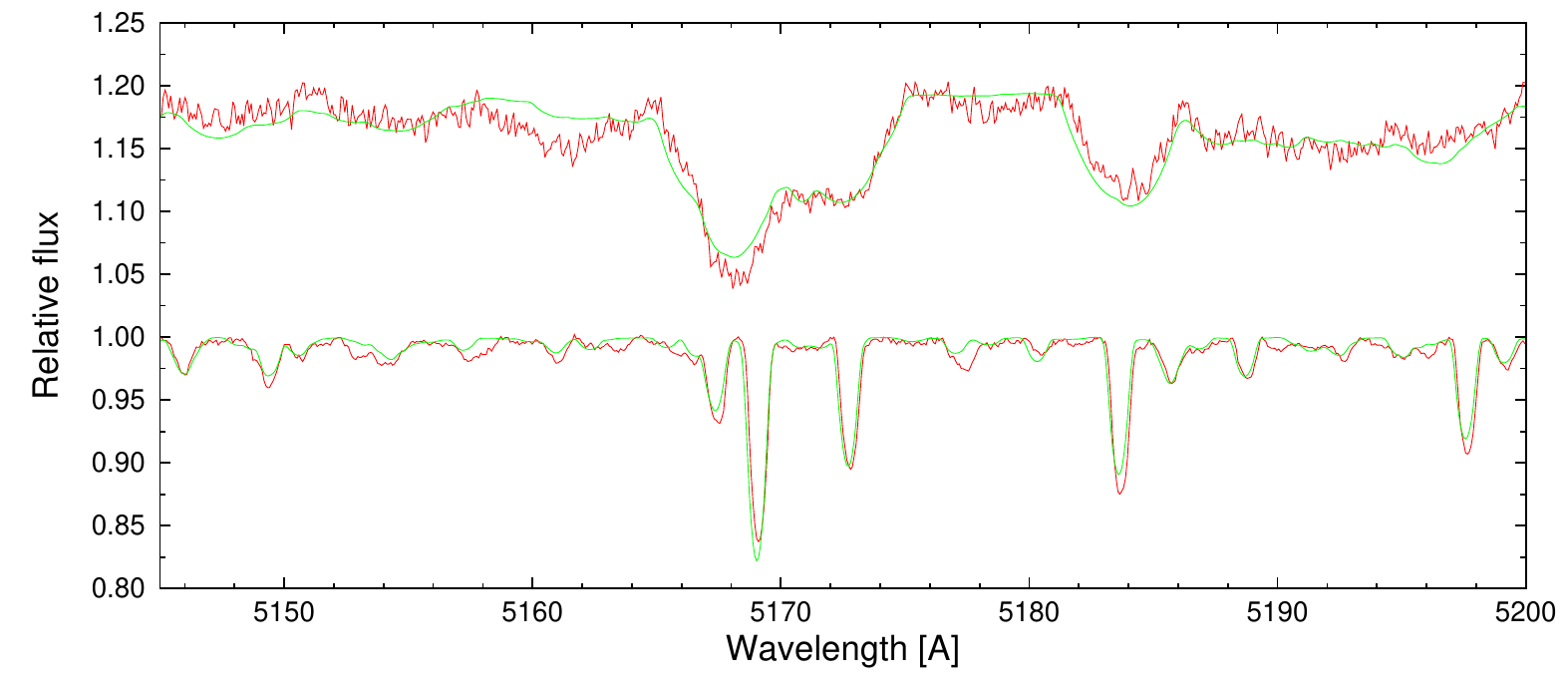}
\end{center}
\caption{Disentangled spectra of the components (red lines) and their best fits with synthetic spectra obtained from iSpec. 
The line depth of the components was corrected for the contribution of the other component. 
The spectrum of the secondary was shifted by +0.2 for clarity}
\label{fig:iSpec}
\end{figure}

The resulting parameters for the primary component in the blue section (4393 - 4639 \AA) of the spectrum 
are $T_{\rm eff}$ = 11\,000$\pm$500 K, $\log g$ = 4.17$\pm$0.45, [M/H] = 0.19$\pm$0.18, 
and $v \sin i$ = 27.9$\pm$2.5 km\,s$^{-1}$. An attempt to adjust the individual abundances of the elements 
did not lead to a significant improvement in the fit. The yellow section of the spectrum (5034 - 5333 \AA) resulted 
in a consistent parameter set: $T_{\rm eff}$ = 11\,000$\pm$400 K, $\log g$ = 4.37$\pm$0.60, [M/H] = 0.35$\pm$0.08, 
and $v \sin i$ = 29.1 $\pm$ 3.2 km\,s$^{-1}$. The surface gravity is consistent with the main-sequence evolutionary status. 
A segment of the yellow part of the spectrum is plotted in Fig.~\ref{fig:iSpec}.

Modeling the secondary component spectrum was difficult and the results are much less robust. 
A comparison of the disentangled spectrum of the secondary component with the synthetic spectra showed that numerous dips 
cannot be identified with the spectral lines but are artifacts resulting from
an incorrect continuum definition. The artifacts are much less numerous in the yellow region (see Fig.~\ref{fig:Rozpletene}). 
To arrive at a useful fit, the continuum artifacts were skipped. Thus, in the yellow
spectrum, the analysis used the three least-affected segments: 5032 - 5200 \AA, 5213 - 5241 \AA, and 5259 - 5332 \AA. 
To avoid non-physical parameters (e.g. zero surface gravity), 
only the surface temperature and the projected rotational velocity were
adjusted, fixing $\log g$ = 4.2 and [M/H] = 0 (solar metallicity). This resulted in $T_{\rm eff}$ = 8\,420$\pm$130 K and $v \sin i$ = 134 $\pm$ 10 km\,s$^{-1}$. 
The blue section (segments 4440 - 4496 \AA, and 4513 - 4589 \AA) was modeled under the same
assumptions, resulting in $T_{\rm eff}$ = 7\,780$\pm$140 K and $v \sin i$ = 121 $\pm$7 km\,s$^{-1}$. 
We also attempted to determine the parameters from the two segments in the blue part separately. 
While the resulting rotational velocities are similar, $v \sin i$ = 124 $\pm$9 km\,s$^{-1}$ and $v \sin i$ = 118$\pm$11 km\,s$^{-1}$, 
the temperatures differ by almost 600 K. Due to the numerous artifacts in the secondary component's spectrum 
in the blue region, we prefer the parameters obtained from the yellow part of the spectrum.

The {\tt iSpec} code searches for the optimal parameters, using the Levenberg-Marquard
algorithm, and the errors are estimated from the co-variance matrix \citep[see][]{blanco14}. 
The errors strongly depend on the supplied 
uncertainties of the data. The error estimates should also be taken with caution because of the problematic continuum 
normalization (especially of the secondary component). A more reliable parameter estimate would require data with 
a better spectrophotometric calibration and higher spectral resolution.

\section{TESS Photometry}
\label{tessphot}
In this study, we also analyzed the photometric data acquired by the \textit{Transiting Exoplanet Survey Satellite} (\textit{TESS}), 
which is open to public access. The satellite, launched in 2018, was designed as an all-sky space survey searching for exoplanets 
orbiting stars brighter than 12 mag \citep{Ricker2014}. The spacecraft consists of four 100-mm telescopes (f/1.4) with four CCD 
cameras, each with a 4 Mpix chip. The combined field of view is 24~$\times$~96 degrees. 
\textit{TESS} uses the broad bandpass filter (600 -- 1000 nm), which is centered on the \textit{I$_{C}$} filter. 
During the primary mission (from July 2018 to July 2020), nearly the whole sky was observed in 26 sectors. 
Each sector is 27.4 days long. Every 30 minutes, the full-frame image (FFI) was obtained. 
For selected targets, the short-cadence (SC) data are available. They are collected every 2 minutes. 
In July 2020, the mission was extended for the following two years and slightly modified. 
The FFI are obtained every 10 minutes \citep{Bell2020}, and the number of targets with SC data
was increased and, for very interesting targets, the data cadence, as short as 20 seconds, was also collected.

HD\,183986 was observed by \textit{TESS} in Sector 14 from July 18 to August 15, 2019. 
However, only FFIs are available. Their cadence (30 minutes) is not sufficient for precise period analysis 
and produces many spurious low-amplitude frequencies. During the extended mission, this target was observed in two 
consecutive sectors (Sector 40 and 41), from June 24 to August 20, 2021. SC data are also available from these sectors. 
A further period analysis was primarily based on the SC data and the FFIs (with improved 10-minute cadence) serve for 
confirmation of our results. We used an open-source package, \texttt{eleanor} \citep{Feinstein2019} to obtain a light 
curve from \textit{TESS} FFIs. This package cuts a small area (20 $\times$ 20 pixels, i.e. 7 $\times$ 7 arcmin) around 
the target and downloads only this part of the FFIs. In the next stage, it automatically selects the optimal shape and size 
of the aperture. The remaining part of the image cut is used to model the background. 
This workflow, and the quality of the final light curve, is very similar to the PDCSAP\_FLUX data of \textit{TESS} SC data. 
Finally, the brightness of the star was transformed from fluxes to \textit{TESS} magnitudes.

\begin{figure}[h]
\begin{center}
\includegraphics[width=\columnwidth]{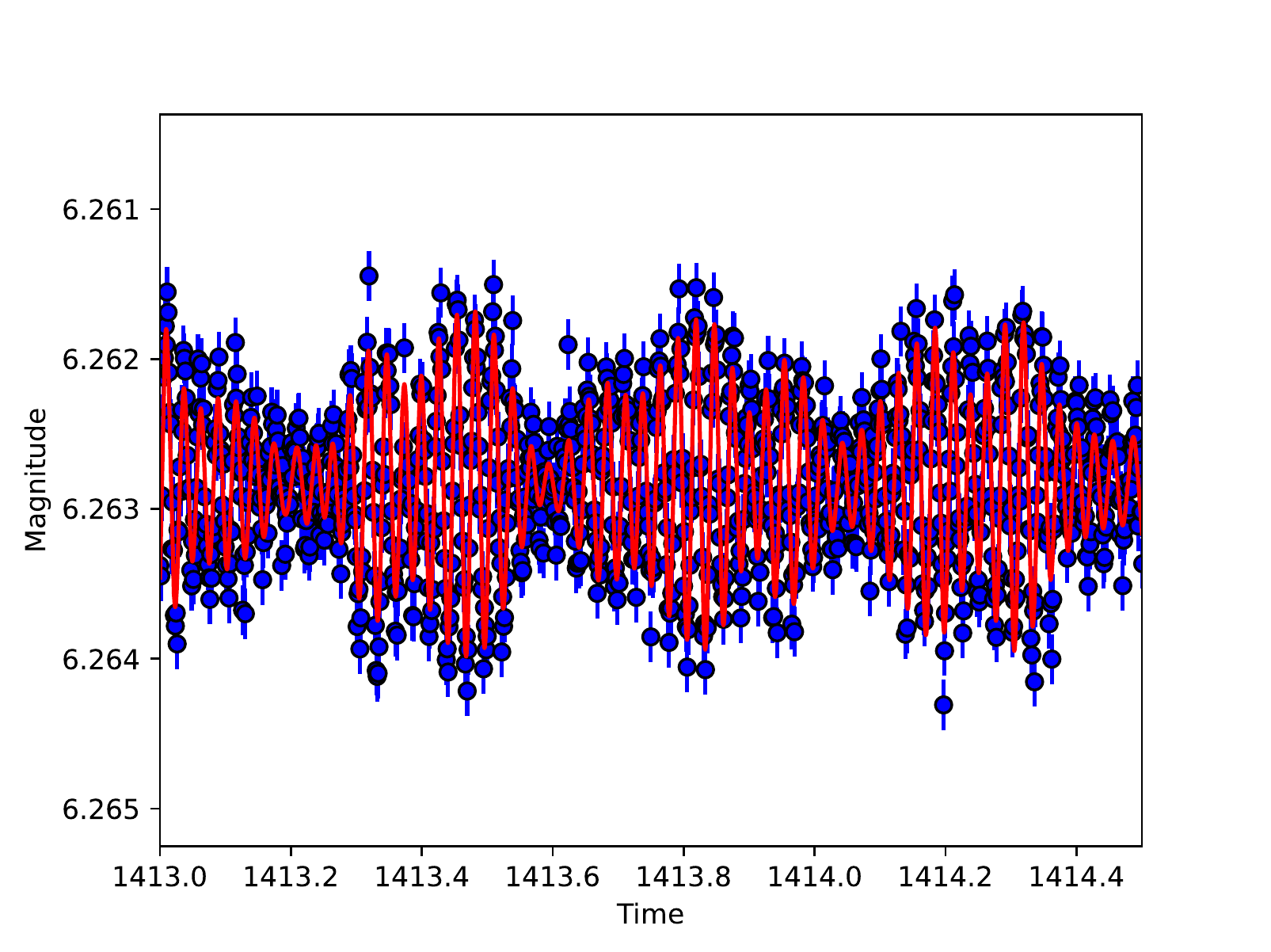}
\end{center}
\caption{Part of light curve of HD\,183986 obtained by \textit{TESS} in SC
mode, together with a model from period analysis. Time is given as BJD - 2\,457\,000.}
\label{fig:model-mag}
\end{figure}

A small part of the light curve of HD\,183986 obtained from \textit{TESS} is shown in Fig. \ref{fig:model-mag}. 
No long-term changes of the light curve could be observed as the observations span a period of about 57 days. 
The pulsations with beats are clearly visible. The period of the beats is about 9.5 hours.
The scatter of data caused by the beats is about 1~mmag around the mean value. The amplitudes of the individual pulsation modes is less than 0.5~mmag.
The mean uncertainty of the data points is about 0.15~mmag. No significant change of brightness between the years 2019 (Sector 14) and 2021 (Sectors 40 and 41) was observed.

\subsection{Period analysis}

\begin{table}[h!]
\caption{Short-periodic signals found in the \textit{TESS} data -- frequencies, periods and amplitudes of associated sine waves. FAP - false alarm probability.}
\label{tab:period}
\begin{center}
{\footnotesize \begin{tabular}{cccc}
	\hline
	      Frequency        &        Period        &      Amplitude      &     FAP      \\
	      (d$^{-1}$)       &        (min)         &       (mmag)        &     (\%)     \\ \hline
	35.80140 $\pm$ 0.00008 & 40.2219 $\pm$ 0.0001 & 0.4670 $\pm$ 0.0026 & $<10^{-322}$ \\
	38.18852 $\pm$ 0.00009 & 37.7077 $\pm$ 0.0001 & 0.3154 $\pm$ 0.0020 & $<10^{-322}$ \\
	29.86847 $\pm$ 0.00016 & 48.2114 $\pm$ 0.0003 & 0.1632 $\pm$ 0.0019 & $<10^{-322}$ \\
	32.26096 $\pm$ 0.00022 & 44.6360 $\pm$ 0.0003 & 0.1083 $\pm$ 0.0018 & $<10^{-322}$ \\
	33.99184 $\pm$ 0.00025 & 42.3631 $\pm$ 0.0003 & 0.0905 $\pm$ 0.0017 & $<10^{-322}$ \\
	36.03922 $\pm$ 0.00035 & 39.9565 $\pm$ 0.0004 & 0.0668 $\pm$ 0.0017 & $<10^{-322}$ \\
	26.76969 $\pm$ 0.00040 & 53.7922 $\pm$ 0.0008 & 0.0576 $\pm$ 0.0017 & $10^{-255}$  \\
	36.18048 $\pm$ 0.00057 & 39.8005 $\pm$ 0.0006 & 0.0416 $\pm$ 0.0016 & $10^{-134}$  \\
	35.10046 $\pm$ 0.00055 & 41.0251 $\pm$ 0.0006 & 0.0403 $\pm$ 0.0016 & $10^{-119}$  \\
	38.20103 $\pm$ 0.00050 & 37.6953 $\pm$ 0.0005 & 0.0358 $\pm$ 0.0016 & $10^{-100}$  \\
	27.10228 $\pm$ 0.00064 & 53.1321 $\pm$ 0.0013 & 0.0344 $\pm$ 0.0016 &  $10^{-94}$  \\
	35.78888 $\pm$ 0.00058 & 40.2360 $\pm$ 0.0006 & 0.0317 $\pm$ 0.0016 &  $10^{-80}$  \\
	38.17779 $\pm$ 0.00083 & 37.7183 $\pm$ 0.0008 & 0.0248 $\pm$ 0.0016 &  $10^{-47}$  \\
	34.13131 $\pm$ 0.00088 & 42.1900 $\pm$ 0.0011 & 0.0239 $\pm$ 0.0016 &  $10^{-43}$  \\
	35.81213 $\pm$ 0.00098 & 40.2098 $\pm$ 0.0011 & 0.0217 $\pm$ 0.0016 &  $10^{-35}$  \\
	38.54971 $\pm$ 0.00116 & 37.3544 $\pm$ 0.0011 & 0.0213 $\pm$ 0.0016 &  $10^{-34}$  \\
	47.20413 $\pm$ 0.00116 & 30.5058 $\pm$ 0.0007 & 0.0194 $\pm$ 0.0016 &  $10^{-27}$  \\
	27.17201 $\pm$ 0.00114 & 52.9957 $\pm$ 0.0022 & 0.0189 $\pm$ 0.0016 &  $10^{-26}$  \\
	31.38478 $\pm$ 0.00127 & 45.8821 $\pm$ 0.0019 & 0.0176 $\pm$ 0.0016 &  $10^{-21}$  \\
	33.96859 $\pm$ 0.00124 & 42.3921 $\pm$ 0.0016 & 0.0154 $\pm$ 0.0016 &  $10^{-15}$  \\
	40.98869 $\pm$ 0.00146 & 35.1317 $\pm$ 0.0013 & 0.0142 $\pm$ 0.0016 &  $10^{-12}$  \\
	43.37044 $\pm$ 0.00183 & 33.2023 $\pm$ 0.0014 & 0.0129 $\pm$ 0.0016 &  $10^{-9}$   \\
	35.52424 $\pm$ 0.00170 & 40.5357 $\pm$ 0.0019 & 0.0124 $\pm$ 0.0016 &  $10^{-8}$   \\
	40.96186 $\pm$ 0.00201 & 35.1546 $\pm$ 0.0017 & 0.0122 $\pm$ 0.0016 &  $10^{-7}$   \\
	27.96951 $\pm$ 0.00179 & 51.4846 $\pm$ 0.0033 & 0.0120 $\pm$ 0.0016 &  $10^{-7}$   \\
	73.98813 $\pm$ 0.00173 & 19.4626 $\pm$ 0.0005 & 0.0119 $\pm$ 0.0016 &  $10^{-7}$   \\
	27.66911 $\pm$ 0.00190 & 52.0436 $\pm$ 0.0036 & 0.0119 $\pm$ 0.0016 &  $10^{-7}$   \\
	37.12996 $\pm$ 0.00207 & 38.7827 $\pm$ 0.0022 & 0.0116 $\pm$ 0.0016 &  $10^{-6}$   \\
	34.47284 $\pm$ 0.00224 & 41.7720 $\pm$ 0.0027 & 0.0110 $\pm$ 0.0015 &  $10^{-5}$   \\
	38.24037 $\pm$ 0.00211 & 37.6565 $\pm$ 0.0021 & 0.0107 $\pm$ 0.0015 &  $10^{-4}$   \\
	38.47998 $\pm$ 0.00241 & 37.4221 $\pm$ 0.0023 & 0.0102 $\pm$ 0.0015 &  $10^{-4}$   \\
	37.41784 $\pm$ 0.00227 & 38.4843 $\pm$ 0.0023 & 0.0093 $\pm$ 0.0015 &    0.019     \\
	26.38882 $\pm$ 0.00213 & 54.5686 $\pm$ 0.0044 & 0.0093 $\pm$ 0.0015 &    0.020     \\
	45.60557 $\pm$ 0.00246 & 31.5751 $\pm$ 0.0017 & 0.0090 $\pm$ 0.0015 &    0.069     \\ \hline
\end{tabular}}
\end{center}
\end{table}

\begin{table}[h!]
\caption{Long-periodic signals found in the \textit{TESS} data. For
description, see Tab. \ref{tab:period}.}
\label{tab:period-long}
\begin{center}
{\footnotesize \begin{tabular}{cccc}
	\hline
	      Frequency       &       Period        &      Amplitude      &    FAP     \\
	     (d$^{-1}$)       &         (d)         &       (mmag)        &    (\%)    \\ \hline
	0.17702 $\pm$ 0.00178 & 5.6490 $\pm$ 0.0569 & 0.0163 $\pm$ 0.0016 & $10^{-18}$ \\
	0.10729 $\pm$ 0.00166 & 9.3208 $\pm$ 0.1445 & 0.0141 $\pm$ 0.0016 & $10^{-11}$ \\
	0.23067 $\pm$ 0.00168 & 4.3353 $\pm$ 0.0316 & 0.0137 $\pm$ 0.0016 & $10^{-11}$ \\
	1.22664 $\pm$ 0.00195 & 0.8152 $\pm$ 0.0013 & 0.0116 $\pm$ 0.0016 & $10^{-6}$  \\
	0.19133 $\pm$ 0.00197 & 5.2267 $\pm$ 0.0537 & 0.0106 $\pm$ 0.0015 & $10^{-4}$  \\
	0.32543 $\pm$ 0.00217 & 3.0728 $\pm$ 0.0205 & 0.0112 $\pm$ 0.0015 & $10^{-6}$  \\
	0.41663 $\pm$ 0.00249 & 2.4002 $\pm$ 0.0143 & 0.0102 $\pm$ 0.0015 & $10^{-3}$  \\
	1.00849 $\pm$ 0.00324 & 0.9916 $\pm$ 0.0032 & 0.0100 $\pm$ 0.0015 &   0.001    \\
	6.86095 $\pm$ 0.00219 & 0.1458 $\pm$ 0.0001 & 0.0098 $\pm$ 0.0015 &   0.003    \\
	0.91730 $\pm$ 0.00214 & 1.0901 $\pm$ 0.0026 & 0.0094 $\pm$ 0.0015 &   0.018    \\
	0.99419 $\pm$ 0.00211 & 1.0058 $\pm$ 0.0022 & 0.0090 $\pm$ 0.0015 &   0.079    \\ \hline
\end{tabular}}
\end{center}
\end{table}

\begin{figure}[h]
\begin{center}
\includegraphics[width=\columnwidth]{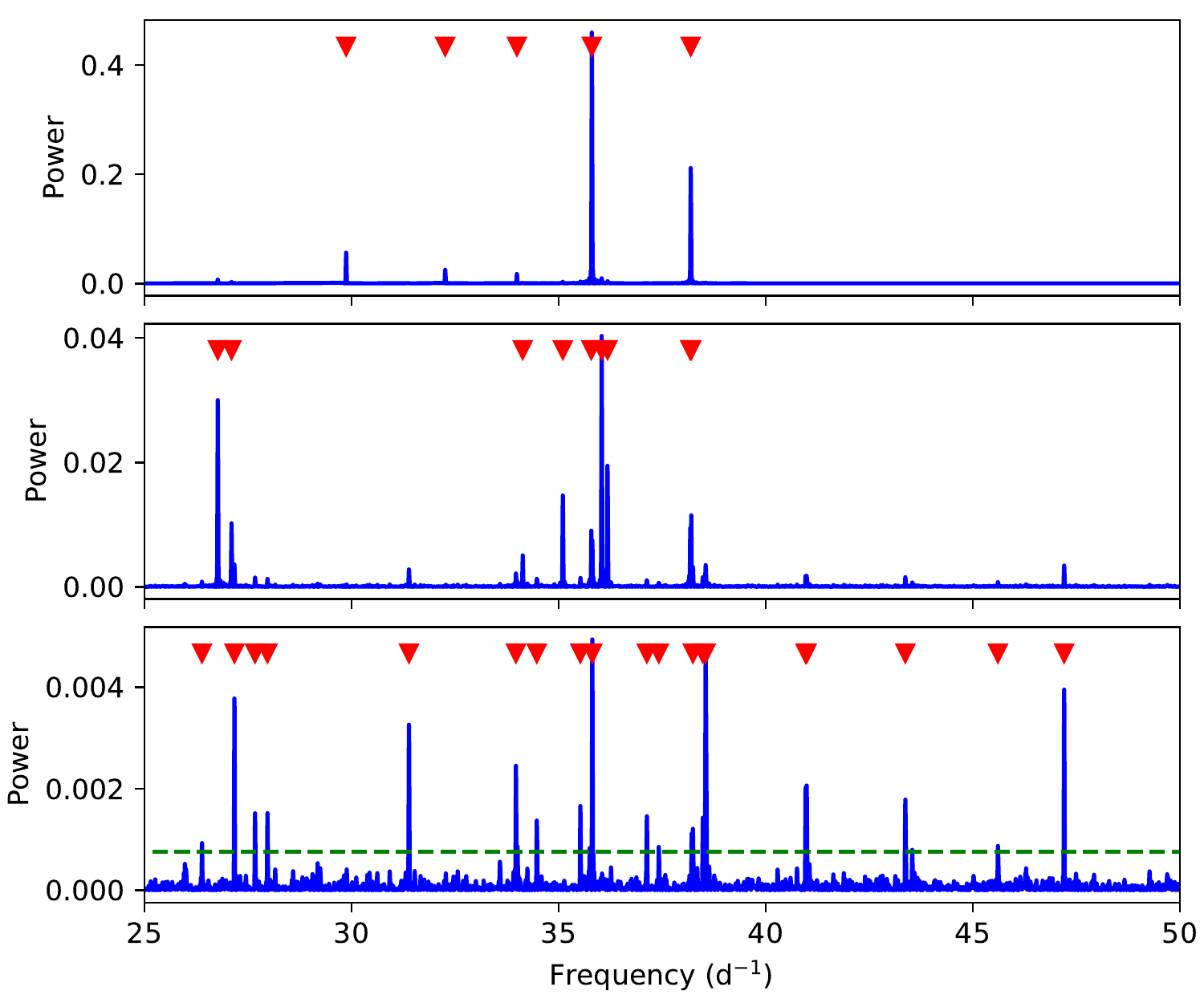}
\includegraphics[width=\columnwidth]{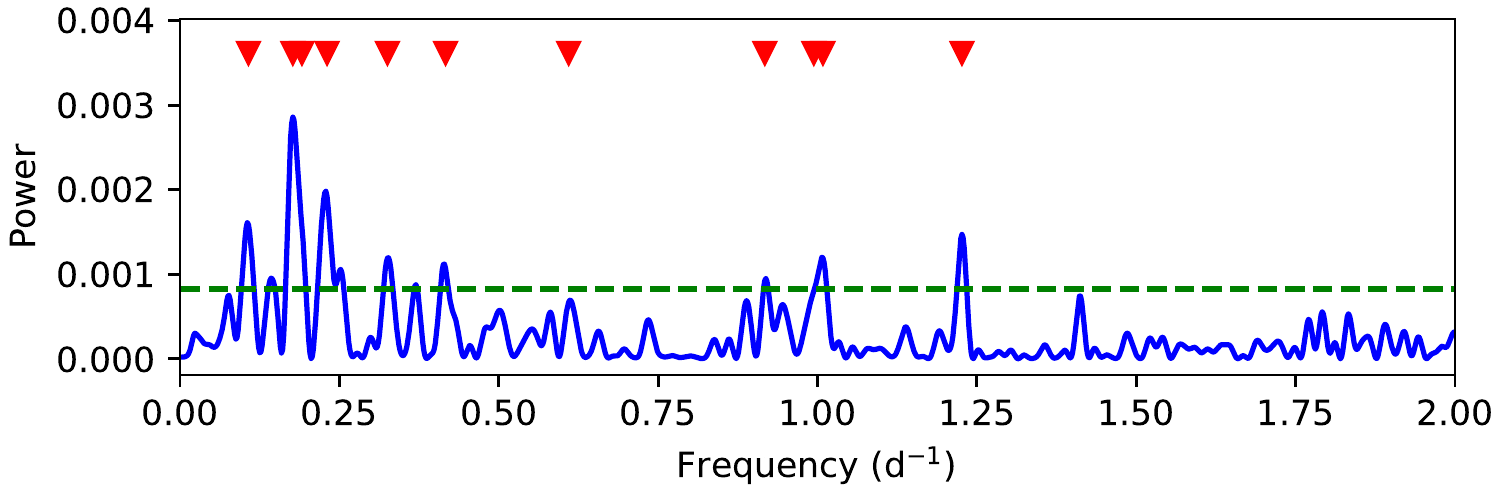}
\end{center}
\caption{Periodograms associated to the SC data. The dashed horizontal line provides FAP (false alarm probability) 
levels of 0.3\%. The red triangles show the frequencies found, and listed in Tab. \ref{tab:period}. 
\textit{Top:} high frequencies (periods less than an hour). \textit{Bottom:} low frequencies (periods more than one day).}
\label{fig:power}
\end{figure}

The Generalized Lomb-Scargle \citep[GLS;][]{Zechmeister2009} periodogram was used to find all
of the frequencies in the data. 
At first, we constructed a GLS periodogram from the original data. We identified the highest power peak and 
subtracted the associated sine wave from the light curve. Next, we generated a GLS periodogram only 
from the obtained residua. We repeated this procedure until the false-alarm probability (FAP) level of 0.3\% was reached. 
This value corresponds to $3\sigma$-confidence. Using this procedure, we removed all aliases.

We found 45 frequencies (listed in Tab. \ref{tab:period} and \ref{tab:period-long}) with FAP
levels less than 0.3\%. The amplitudes of periodic signals with
a FAP above this limit are significantly smaller than the precision of the input photometric data. 
Therefore, we assumed that they are only noise. The corresponding periods 
of the detected signals were in a range from 30 minutes to 5.6 days. 
Most of them are shorter than one hour. The amplitudes of the associated 
sine waves were smaller and range from 0.5 mmag to less than 0.01 mmag (for the most powerless frequencies). 
Periodograms, where all the frequencies are marked, are in Fig. \ref{fig:power}. The general shape of the light curve 
is given by the 5 most significant frequencies with amplitudes larger than approx. 0.1~mmag (top panel of Fig. \ref{fig:power}; 
with periods 40.2, 37.7, 48.2, 44.6 and 42.4 minutes). The reality of some frequencies (mainly those in the third
panel of Fig. \ref{fig:power}) with small amplitudes and a relatively high FAP is questionable. 
Their amplitudes are on a level of errors of \textit{TESS} observations and, therefore, they could only
be the result of noise in the light curve. 
Photometric data with a higher precision, obtained over a longer time interval, are needed to confirm these frequencies.
On the other side of our periodograms, we found 11 signals with longer periods (from a few hours to a few days; 
bottom panel of Fig. \ref{fig:power} and Tab. \ref{tab:period-long}). 
These frequencies probably result from instrumental effects.

We used photometric data obtained from FFI, with a cumulative exposure time of 10 minutes, to check the results 
of our period analysis. The main advantage of using this data is that the photometric precision is more than as twice good as 
the SC data -- the mean uncertainty is 0.068 mmag. 
Their time resolution is also sufficient to distinguish 40-minutes pulsations. 
The Nyquist frequency is 72~d$^{-1}$, with a corresponding period of 20 minutes. We found 36 frequencies with a FAP below our 
$3\sigma$ level. All the most significant frequencies found using the SC data were confirmed. 
However, only one of the low frequencies (with a period of 0.8152 days or 19.6 hours) is
also present in FFI data. 
This fact confirms our previous hypothesis that all long-periodic signals are spurious. 
Moreover, the amplitude of this single long-periodic signal present in both data sets is significantly 
below the precision of the input data. Therefore, its nature and reality is more than questionable.

\subsection{A $\delta$~Scuti pulsations}

In general, photometric variations with amplitudes 0.003 to 0.9 mag, with periods of 0.02 d - 0.25 d, 
in stars within the spectral type range A0 to F5, are typical of $\delta$~Scuti stars
\citep{Breger2000a}. 
Since the majority of $\delta$~Scuti stars are multi-mode pulsators, the most plausible explanation
is that the pulsator could be the secondary component of the binary HD\,183986. According to \cite{Fernie1964}, 
the relation between the period and the mass and radius of a radially pulsating star is

\begin{equation}
P=Q\times(M_{\star}/M_{\sun})^{-1/2}\times
(R_{\star}/R_{\sun})^{3/2}.
\end{equation}

\citet*{Breger1975} deduced average values $Q=0.033, 0.025$ and 0.020 days for the fundamental frequency, 
its first and second overtones. The values may vary due to, for example, rotation, or differ for hot and 
cool $\delta$ Scuti-type stars. 

With the mass and radius of the secondary estimated in Sec. \ref{iSpec}, we get a fundamental frequency 
of $f_{0}=19.8$\, d$^{-1}$\, and with the values for the primary, we get $f_{0}=12.07$\, d$^{-1}$. Neither of the 
frequencies was detected in the TESS photometry. The observed frequencies are, therefore, higher non-radial pulsational modes.

It is important to note that mode identification is more difficult for $\delta$ Scuti stars because they are located 
at the intersection between the classical instability strip and the main sequence in the HR diagram, a region where 
the asymptotic theory of non-radial pulsations is invalid due to low-order modes (complicated by avoided crossing and
mixed modes). Although there are examples of mode identification based on long-term ground-based observations 
(e.g. $\tau$~Peg, \citet{Kennelly1998} or 4~CVn, \citet{Breger2000b}), only a few modes have been identified by comparing 
the observed frequencies with modelling.    

Concerning the low-amplitude $\delta$ Scuti stars (LADS), HD\,174936 \citep{Garcia2009}, HD\,50844 \citep{Poretti2009}, 
and HD\,50870 \citep{Mantegazza2012} showed an extremely rich frequency content of low-amplitude peaks in the range 
0-35 d$^{-1}$\,. A similar dense distribution was obtained in KIC\,4840675
\citep{Balona2012a}. However, KIC\,9700322 \citep{Breger2011}, one of the coolest $\delta$ Scuti stars with 
$T_{\rm eff}$ = 6\,700~K, revealed a remarkably simple frequency content, with only two radial modes and a large 
number of combination frequencies and rotational modulations. Based on the MOST satellite data, \citet{Monnier2010} identified 
57 distinct pulsation modes in $\alpha$ Oph above a stochastic granulation noise.

A good example of a high-amplitude $\delta$ Scuti (HADS) star is V2367~Cyg. Almost all the light variation of V2367 Cyg 
\citep{Balona2012b} is attributed to three modes and their combination frequencies. 
The authors also detected several hundred other frequencies of very low amplitude in the star 
(with $T_{\rm eff}$ = 7\,300~K). On the other hand, twelve independent terms, beside the radial fundamental mode and its 
harmonics up to the tenth harmonic, were identified in the light of CoRoT 101155310 \citep{Poretti2011}. 
Regarding the linear combinations of modes, only 61 frequencies were found, down to 0.1 mmag. 
A much smaller number of low-amplitude modes were thus reported for this HADS star, although it has the 
same effective temperature as V2367 Cyg. As the examples show, a large number of low-amplitude modes
have been detected in both of the largest subgroups of $\delta$ Scuti stars by various space
missions. 

Although an investigation of the stellar energy balance proved that $\delta$ Scuti stars are energetically and mechanically 
stable even when hundreds of pulsational modes are present \citep{Moya2010}, the number of modes can also be interpreted as 
non-radial pulsation superimposed on granulation noise \citep{Kallinger2010}.

It seems that in HD\,183986, we have reached a level of precision where the interpretation of the periodicities arising from 
different physical processes can be quite difficult. The presence of non-radial modes and granulation noise seem to wash out the
physical separation of the LADS and HADS groups. It may be that only the selection mechanism of the excited modes is
different in the two groups \citep{Paparo2019}. However, we do not exactly know the nature of the selection mechanism. 
Any step towards understanding the selection mechanism of non-radial modes in $\delta$ Scuti stars
would, therefore, be very valuable. A meaningful direction would be to find some regularities, if there are any, 
among the increased number of observed frequencies based on a promising parameter such as the frequency spacings.


\section{Evolutionary status of the binary system}
\label{evolution}

An important constraint on the system can be based on stellar evolutionary models. We used a calibration of
the Geneva evolutionary models of non-rotating stars by \cite{lejeune01}. Grids for $Z$ = 0.020 were assumed. We began with an apparent 
$V$ band magnitude and parallax, which are reliable quantities measured to a very good precision. From
them, one can calculate 
the absolute $V$ band magnitude. Assuming a temperature of the primary of 11\,000 K, from {\tt iSpec}
modeling, one can plot the location of the star in the HR diagram. This, in fact, represents an upper limit on the mass 
and brightness of the primary since we did not take into account the contribution of the secondary star. 
However, the secondary is much fainter and 
it will not significantly affect the location of the primary star. Interpolating in the evolutionary tracks, the location of 
the primary corresponds to a star with a mass of 3.2 $M_{\odot}$ and an age of $220\pm 50$ Myr. 
Note that the isochrone is almost isothermal in this region so its error is mainly given by the error in the effective 
temperature of the primary star and is not very sensitive to the brightness or flux dilution due to the secondary star.

\begin{figure}[ht]
\begin{center}
\includegraphics[width=\columnwidth]{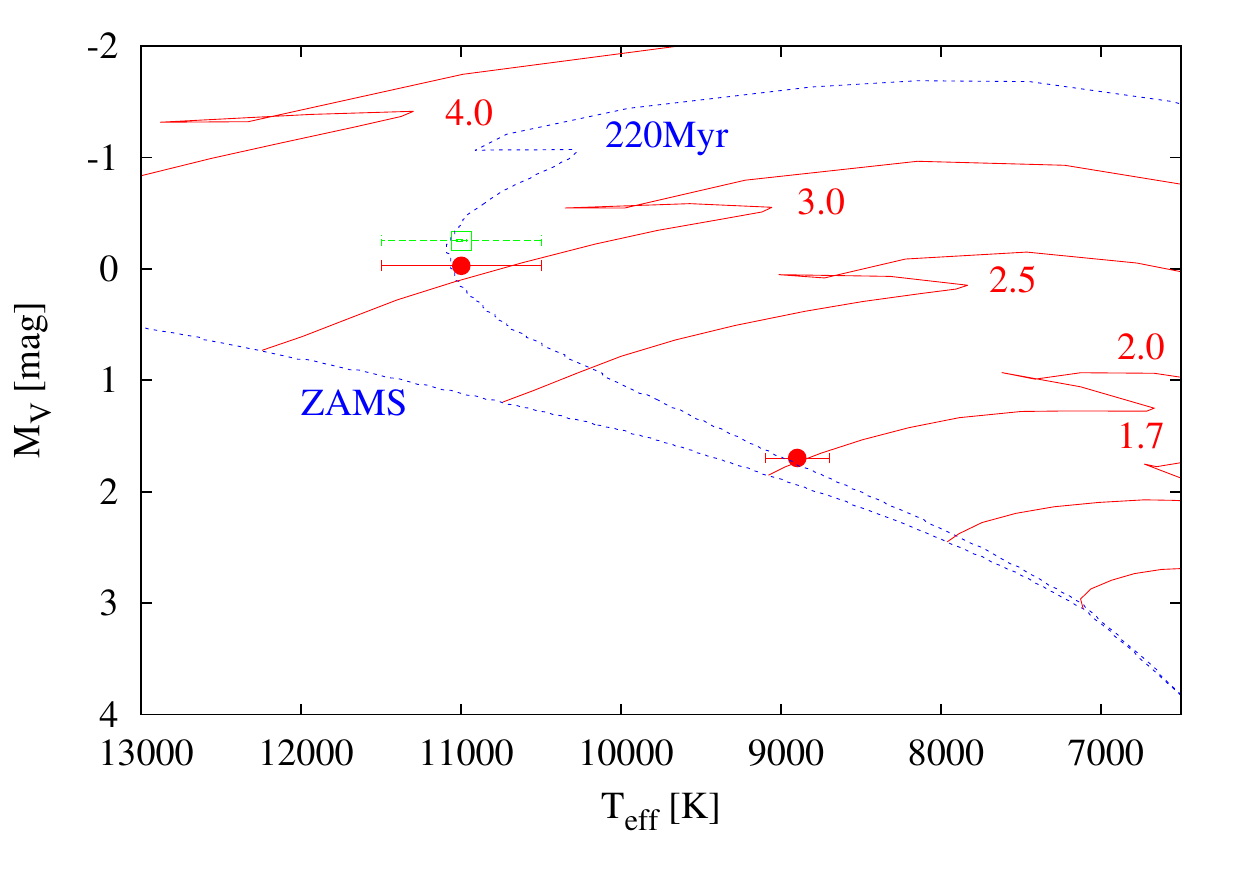}
\end{center}
\caption{Location of the primary and secondary stars in the HR diagram (full red circles). 
The location of the whole binary is a green open square. Red full lines are evolutionary tracks of stars 
with different masses also plotted in red. The two blue dotted lines are isochrones corresponding to ZAMS and 220 Myr.}
\label{fig:evolution}
\end{figure}

Next we exploit the mass ratio of $M_2/M_1$ = 0.655, obtained from the {\tt KOREL}
disentangling, which is also a reliable quantity, and obtain the mass of the secondary of 2.1 $M_\odot$. 
We again interpolate in the evolutionary tracks to obtain the evolution of such a star. 
Assuming that both stars have the same age, the intersection of this track 
with the isochrone of the primary gives us the location of the secondary star in the HR diagram. 
Once we know the brightness of the secondary star, we subtract its value and correct the location of the primary 
and iterate the whole procedure again. The final location of both stars, as well as that of the whole binary, is shown 
in Figure \ref{fig:evolution}. The parameters of the stars are summarized in Table~\ref{tab:evolution}. 
One can see that both stars are located on the main sequence. The masses of the stars from evolutionary models 
are in good agreement with the masses obtained from the spectroscopic orbit. They indicate that the inclination 
of the orbit must be close to 90 degrees.

\begin{table}
\caption{Stellar parameters based on evolutionary models. }
\label{tab:evolution}
\centering
\begin{tabular}{lr}
	\hline\hline
	$T_1$ [K]         & 11000$\pm$ 500 \\
	$M_1$ [M$_\odot$] &   3.1$\pm$ 0.2 \\
	$T_2$ [K]         &  8900$\pm$ 200 \\
	$M_2$ [M$_\odot$] &   2.0$\pm$ 0.2 \\
	Age [Myr]         &    220$\pm$ 50 \\
	$L_2/L$           &           0.19 \\
	$L_1/L$           &           0.81 \\
	$L_2/L_1$         &           0.23 \\ \hline
\end{tabular}
\end{table}

\section{Discussion and conclusions}
\label{concl}

New and extensive \'echelle spectroscopy conclusively showed that HD\,183986 is a spectroscopic binary and led to the first 
reliable determination of the orbital period of the system, $P$ = 1268.2$\pm$1.1 days. 
While the SB1 orbit of the primary component is very robust and well defined, the secondary component is significantly 
fainter and rotating much faster, which complicates the analysis. The lines of the secondary component were, however, clearly 
detected in the wings of the strongest metallic lines, e.g. \ion{Mg}{ii} $\lambda$ 4481\AA~, and in the disentangled spectrum.

To arrive at the SB2 orbital parameters, the Fourier domain disentangling using code {\tt KOREL} was used. 
The disentangling  provided not only the individual component spectra but also the SB2 orbit for the
system, including the mass ratio of the components $M_2/M_1 = 0.655$. Using the observed color
indices, the spectral type of the primary component was estimated as B9.5V. Knowing the mass ratio from the Fourier disentangling 
indicates an A5V secondary star and the flux ratio in the visual region $F_2/F_1$ = 0.245. 
After correcting the spectra for the light contribution of the other component, the disentangled spectra were further 
modeled to determine the atmospheric parameters, $T_{\rm eff}$, $\log g$, [M/H] and the projected rotational velocity 
$v \sin i$, using code {\tt iSpec}. The resulting parameters are consistent with those estimated from the Str\"omgren $uvby\beta$ color indices. 
The {\tt iSpec} modelling proved that the secondary component is a fast rotator.

HD\,183986 is an object which, due to the rapid rotation of the secondary
component, shows a smaller depth of the spectral lines for its temperature and metallicity from the photometric color indices. 
Without the inclination angle of the orbit, we cannot reliably determine the absolute parameters of the components. 
It is, however, highly probable that the next Gaia data release (DR3) in 2022 will provide the visual orbit of the 
system photocenter and the missing inclination angle required to arrive at the true masses. Without eclipses the component 
radii cannot be directly determined.

The components of HD\,183986 show significantly different projected rotational velocities. 
The small rotational velocity of the primary, ${v \sin i}$ = 27 km\,s$^{-1}$, and the large 
rotational velocity of the secondary, ${v \sin i}$ = 120 km\,s$^{-1}$, match perfectly the maxima of the bi-modal 
distribution of rotational velocities found for normal late-B to mid A-type stars \citep{royer04}. 
The true rotational velocities are, however, unknown and due to the long orbital period of
HD\,183986, the spin obliquities of the components may be significantly different.

Substantial information on HD\,183986 was also obtained from high-precision photometry with the \textit{TESS}. 
The satellite provided two month-long and uninterrupted photometry of the object. 
The data show several signals with periods of about 40 minutes and amplitudes less than 0.5 mmag.
Here, we concluded that the observed frequencies are higher non-radial pulsational modes.
Also, some lower frequencies (with few-days periods) were found. However, their nature is probably only the result 
of instrumental effects. A detailed analysis of the observed frequencies is, however, beyond the scope of this paper.

\begin{acknowledgements}
We thank the anonymous referee for his/her useful comments, which improved the quality of the paper.
The authors thank Prof.~Harmanec for his valuable advice, and to V.~Koll\'ar, A. Maliuk, P.~Sivani\v{c} and S. Yu.~Shugarov 
for their observations and technical assistance. This work was supported by the Slovak Research and Development Agency under 
contract No.~APVV-20-0148. This work was also supported by the VEGA grant of the Slovak Academy of Sciences 
No.~2/0031/22. The research of PG was supported by internal grant VVGS-PF-2021-2087 of the 
Faculty of Science, P. J. \v{S}af\'{a}rik University in Ko\v{s}ice. EP acknowledges support of the Erasmus+ program of the 
European Union under grant number 2020-1-CZ01-KA203-078200. ZG was supported by the Hungarian National Research, Development 
and Innovation Office (NKFIH) grant K-125015, the PRODEX Experiment Agreement No. 4000137122 between the ELTE University and 
the European Space Agency (ESA-D/SCI-LE-2021-0025), and by the City of Szombathely under agreement No. 67.177-21/2016.
\end{acknowledgements}

\facility{\textit{TESS}}
\software{FOTEL \citep{Hadrava1990}, KOREL \citep{Hadrava2004}, iSpec \citep{blanco14,blanco19}, 
eleanor \citep{Feinstein2019}, IRAF \citep{Tody1986,Tody1993}, PyAstronomy \citep{pya}}

\bibliography{bibfile}{}
\bibliographystyle{aasjournal}

\end{document}